\documentclass{PoS}
\usepackage{amsmath}
\usepackage{xspace}
\usepackage{subfig}
\usepackage{atlasunit}
\usepackage{atlasmisc}
\usepackage{atlasjournal}
\usepackage{DiphotonPublication-defs}

\title{Search for a high mass diphoton resonance using the ATLAS detector}

\ShortTitle{Search for a high mass diphoton resonance using the ATLAS detector}

\author{\speaker{Bruno Lenzi}\\%
      \textit{on behalf of the ATLAS Collaboration}\\
      CERN\\
      E-mail: \email{bruno.lenzi@cern.ch}}

\abstract{A search for new spin-0 resonances decaying into two photons in the ATLAS experiment at the
LHC is described. The analysis is based on $pp$ collision data at $\sqrt{s}$=13~\TeV\ corresponding to integrated luminosities of \intLumiFifteen\ and \intLumiSixteen\ 
recorded in 2015 and 2016, respectively.
A deviation from the Standard Model background-only hypothesis corresponding
to \localSigSpinZeroFifteen\ 
standard deviations is observed in the 2015 data for
a resonance mass hypothesis of \bestMassSpinZeroFifteen~\GeV.
No significant excess at such mass over the background expectation
is observed in the 2016 data.
Limits on the production cross section times branching ratio to two photons
of such resonances are reported.}

\FullConference{38th International Conference on High Energy Physics\\
		  3-10 August 2016\\
		 Chicago, USA}

\begin{document}

\section{Introduction}
\label{sec:intro}

New high-mass states decaying into two photons are predicted in many 
extensions of the Standard Model (SM).
The diphoton final state provides a clean experimental signature with excellent invariant mass resolution
and moderate backgrounds.

Using 3.2--3.3~\ifb\ of $\sqrt{s}=13$~\TeV\ proton--proton ($pp$) collision data recorded in 2015 at the CERN Large Hadron Collider (LHC), the ATLAS and CMS Collaborations
reported an excess in the diphoton invariant mass spectra with respect to the SM continuum background near
the mass value of 750~\GeV~\cite{Aaboud:2016tru,Khachatryan:2016hje}.
The searches were performed using two benchmark signal models, the lightest Kaluza--Klein~\cite{ModernKK} spin-2
graviton excitation ($G^*$) of a Randall--Sundrum~(RS)~\cite{Randall:1999ee} model
or a spin-0 resonance ($X$).
The ATLAS results in Ref.~\cite{Aaboud:2016tru} correspond to a global significance of about
two standard deviations. 

An update of this search is given in Ref.~\cite{ATLAS-CONF-2016-059} and summarised below. The analysis follows closely the description given in Ref.~\cite{Aaboud:2016tru}, but limited to the spin-0 resonance search analysis. Both the 2015 and 2016 $pp$ collision datasets are used, 
corresponding to a total integrated luminosity of \intLumi.

\section{Description of the analysis}
\label{sec:analysis}

The search uses events from $pp$ collisions recorded by the ATLAS detector~\cite{PERF-2007-01} using a diphoton trigger with a signal efficiency close to 99\% for events fulfilling the final event selection. Photon candidates, reconstructed from clusters of energy deposited in the electromagnetic calorimeter and tracks and conversion vertices reconstructed in the inner detector, are required to fullfil tight identification criteria based primarily on shower shapes in the calorimeter. To further reject the background from jets misidentified as photons, the photon candidates are required to be isolated using both calorimeter and tracking detector information.

The transverse energy is required to be $\et > 0.4 \cdot m_{\gamma\gamma}$ for the photon with the highest \et\
and $\et > 0.3 \cdot m_{\gamma\gamma}$ for the photon with the second-highest $\et$,
for a given value of the diphoton invariant mass $m_{\gamma\gamma}$, thus selecting events in which the photons are
preferentially emitted in the central part of the detector. Only events with $m_{\gamma\gamma} >150$~\GeV\ are retained.
With these requirements, \yieldSpinZero~events are selected in the data. The selected sample mainly consists of events 
from diphoton production, with an estimated purity of $(90^{+3}_{-10})$\%. 


Simulated Monte Carlo (MC) samples are used to optimize the search strategy and to study background sources. The invariant mass distribution of the diphoton pair for the signal is expected to peak near the
assumed mass of the new particle, with a spread given by the convolution of its intrinsic decay width
with the experimental resolution, which varies from 2.3~\GeV\
at a mass of 200~\GeV\ to 15~\GeV\ at a mass of 2~\TeV. A double-sided
Crystal Ball (DSCB) function, with a Gaussian core and power-law functions describing the low and high mass sidebands is used to model the experimental resolution of the reconstructed invariant mass. The parameters of the DSCB function are obtained from fits to the invariant mass distributions of simulated narrow-width signal samples. The signal mass distribution for any value of the mass and width hypothesis
is obtained by a convolution of the intrinsic detector resolution with the predicted line-shape distribution of the resonance that combines a Breit-Wigner function, the parton luminosity and the matrix element, calculated using an effective field theory approach at next-to-leading order in QCD.

The estimate of the background $m_{\gamma\gamma}$ contribution in the selected sample
is based on a fit using the following functional form,
with parameters determined from the data:

\begin{equation}
  f_{(k)}(x;b,\{a_{k}\}) = N (1 - x^{1/3})^b x^{\sum_{j=0}^k a_j (\log x)^j} \, ,
  \label{eq:bck_func}
\end{equation}
where $x = m_{\gamma\gamma}/\sqrt{s}$, 
$b$ and $a_{k}$ are free parameters, and $N$ is a normalization factor. The mass distribution from data is fitted in the range above 180~\GeV, and
the search range for the signal is 200--2400~\GeV. To validate the choice of this functional form and to derive the corresponding uncertainties, the method
detailed in Ref.~\cite{HIGG-2013-08} is used to check that the form is flexible enough to accommodate different
physics-motivated underlying distributions. The simplest choice $k=0$ is adopted. The bias related to the choice of functional form for a narrow-signal hypothesis varies from 18 events at 200~\GeV\ to 0.012 events at 2400~\GeV. For larger hypothesized signal widths, the signal is integrated
over a wider mass range and the background uncertainty is larger, varying from 117 events at 200~\GeV\ 
to 0.35 events at 2400~\GeV,
for a hypothesized signal with a relative width $\Gamma_X/m_X$ of 10\%.

The numbers of estimated signal and background events are obtained from maximum-likelihood fits of the $m_{\gamma\gamma}$
distribution of the selected events. The function used to describe the data can be written as
\begin{equation}
N_{\mathrm{S}}(\sigma_\mathrm{S}) f_{\mathrm{S}} (m_{\gamma\gamma}) + N_{\mathrm{B}} f_{\mathrm{B}} (m_{\gamma\gamma}),
\end{equation}
where $N_{\mathrm{S}}$ is the fitted number of signal events, $f_{\mathrm{S}} (m_{\gamma\gamma})$ 
is the normalized invariant mass distribution for a given
signal hypothesis, $N_{\mathrm{B}}$ is the fitted number of background events 
and $f_{\mathrm{B}} (m_{\gamma\gamma})$ is the normalized invariant mass distribution
of the background events. 
The fitted number of signal events is related to the assumed signal cross section times branching ratio to two photons ($\sigma_S$)
in the fiducial acceptance via the integrated luminosity and the
total efficiencies of the event reconstruction, identification and isolation criteria, which ranges from 66\% for a particle with a mass of 200~\GeV\ to
74\% for a mass of 700~\GeV\ and is almost constant above 700~\GeV.

Uncertainties in the signal parameterization, in the
detector efficiency correction factors for the signal 
and in the signal extraction are included in the fit via nuisance parameters, constrained with Gaussian or log-normal penalty terms. The compatibility with the background-only hypothesis when testing a given
signal hypothesis ($m_X$, $\alpha$) is estimated using the local $p$-value ($p_0$) 
based on the profile likelihood ratio test statistic under the asymptotic approximation. Global significance values are computed to account for the trial factors given by the search range with a large number of background-only pseudo-experiments. The expected and observed 95\% confidence level (CL) exclusion limits
on the cross section times branching ratio to two photons are computed using 
a modified frequentist approach.

\section{Results and conclusion}
\label{sec:results}

The 2015 data and simulated samples, used in Ref.~\cite{Aaboud:2016tru}, have been reprocessed with the same reconstruction
software as used for the 2016 data processing, 
which includes small improvements in the reconstruction and selection of converted photons and in the photon energy calibration. Figure~\ref{fig:mgg_distributions_spin0} shows the diphoton invariant mass distribution
together with the background-only fit, for events selected in the full dataset.

\begin{figure}[!h]
\begin{center}
\includegraphics[width=.5\textwidth]{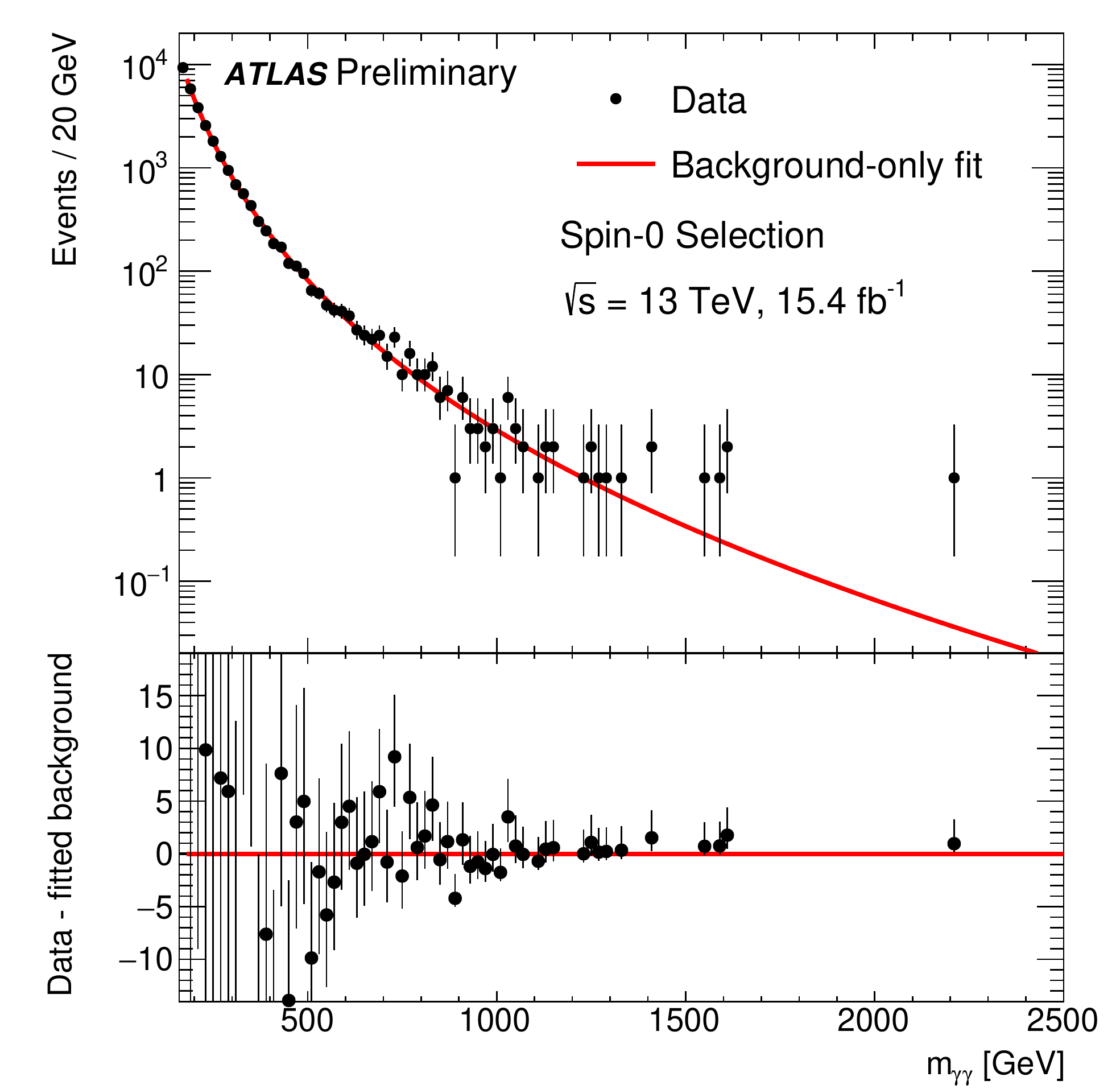}
\end{center}
\caption{Distribution of the diphoton invariant mass of the selected events, together with
  the background-only fit. The difference between the data and this fit is shown in the bottom panel.
  The arrow shown in the lower panel indicates a values outside the range with more than one standard deviation.
  There is no data event with  $m_{\gamma\gamma}>2500$~\GeV. Extracted from Ref.~\cite{ATLAS-CONF-2016-059}.}
\label{fig:mgg_distributions_spin0}
\end{figure}

\begin{figure}[!h]
    \begin{center}
    \subfloat[][]{\label{fig:p0_2D_spin0}\includegraphics[width=.45\textwidth]{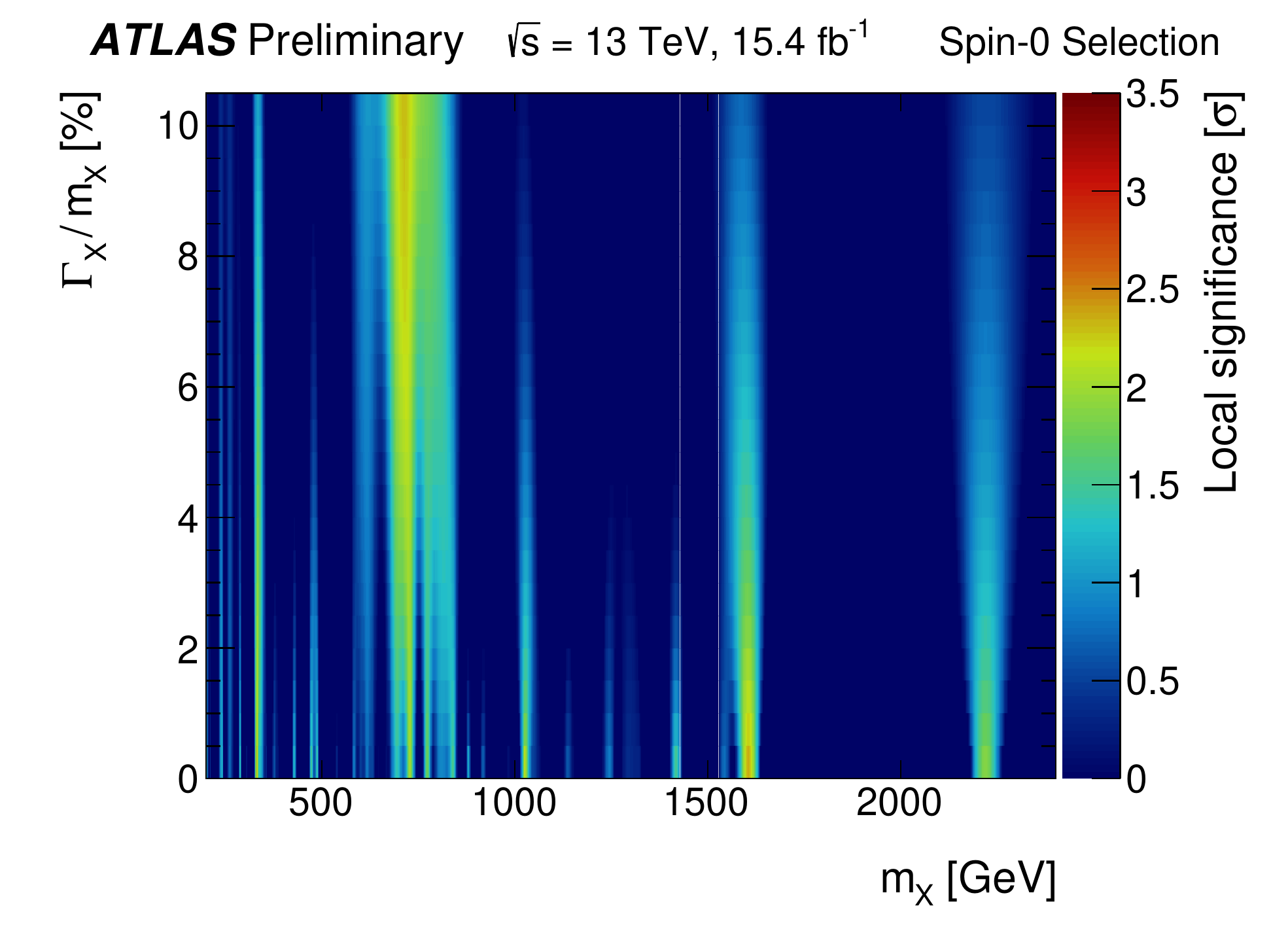}}
    \subfloat[][]{\label{fig:spin0_limit}\includegraphics[width=.45\textwidth]{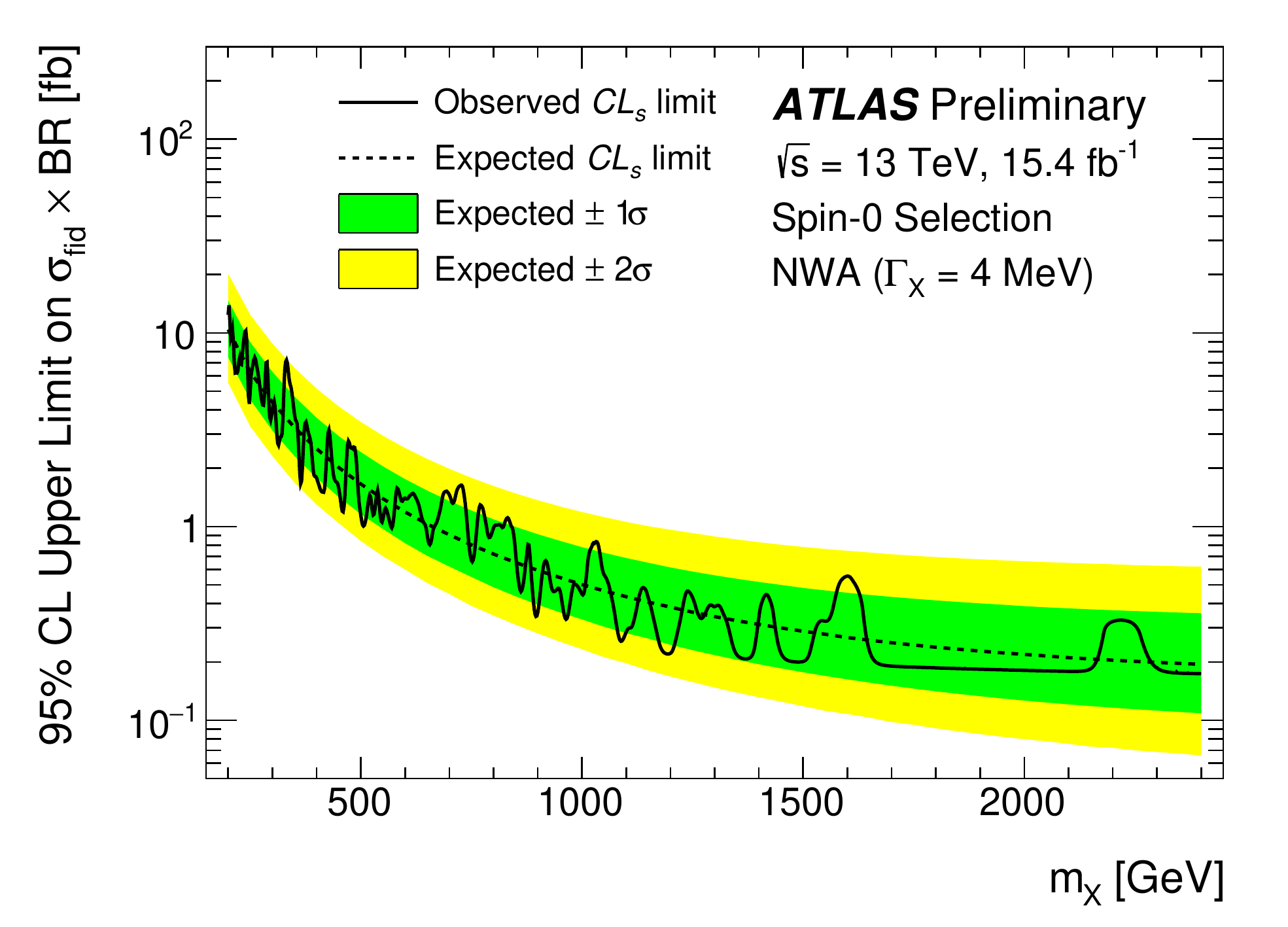}}
    \end{center}
   \caption{(a) Compatibility, in terms of local significance $\sigma$,
     with the background-only hypothesis as a function of the assumed
     signal mass and relative width for a spin-0 resonance. (b) Upper limits on the fiducial cross section times branching ratio to two photons at $\sqrt{s} = 13$~\TeV\ of a spin-0 particle as a function of its mass $m_X$ for a narrow-width signal (NWA) with $\Gamma_X$ = 4~\MeV. Both figures were extracted from Ref.~\cite{ATLAS-CONF-2016-059}.}
    \end{figure}

\newpage
The compatibility with the background-only hypothesis
is computed as a function of the hypothesized resonance mass and width, as
shown in Figure~\ref{fig:p0_2D_spin0}. The largest deviation over the background-only
hypothesis is observed at a mass of \bestMassSpinZero~\GeV\ for an assumed narrow width, corresponding to a local significance 
of \localSigSpinZero\ standard deviations.
In the 700--800~\GeV\ mass range, where the largest deviation from the background-only
hypothesis is observed in the reprocessed 2015 dataset, the largest local significance is 2.3 standard deviations
for a mass near 710~\GeV\ and a relative width of 10\%. 
The global significance of these excesses is less than one standard deviation.

Figure~\ref{fig:spin0_limit} shows the limits on the signal fiducial cross section times branching ratio to two photons for a spin-0 particle as a function of the assumed signal mass for a narrow-width signal.

\bibliographystyle{JHEP}
\bibliography{DiphotonPublication}

\end{document}